\documentclass[aps,prc,twocolumn,superscriptaddress]{revtex4-1}
\usepackage{amssymb}
\usepackage{graphicx}
\usepackage{wasysym}
\usepackage{verbatim}
\usepackage{booktabs}

\usepackage{multirow}
\usepackage{dcolumn}
\newcolumntype{y}{D{-}{-}{3.3}}
\usepackage[bookmarks, hidelinks]{hyperref}

\begin{document}

\title{Performance of Geant4 in simulating semiconductor particle detector response in the energy range below 1 MeV}

\author{G. Soti}
\email[Corresponding author: ]{gergelj.soti@fys.kuleuven.be}
\affiliation{K.U.Leuven, Instituut voor Kern- en Stralingsfysica, Celestijnenlaan 200D, B-3001 Leuven, Belgium}%
\author{F. Wauters}
%\affiliation{K.U.Leuven, Instituut voor Kern- en Stralingsfysica, Celestijnenlaan 200D, B-3001 Leuven, Belgium}%
\altaffiliation[Current address: ]{Center for Experimental Nuclear Physics and Astrophysics, University of Washington, Seattle, Washington 98195, USA}
\author{M. Breitenfeldt}
\affiliation{K.U.Leuven, Instituut voor Kern- en Stralingsfysica, Celestijnenlaan 200D, B-3001 Leuven, Belgium}%
\author{P. Finlay}
\affiliation{K.U.Leuven, Instituut voor Kern- en Stralingsfysica, Celestijnenlaan 200D, B-3001 Leuven, Belgium}%
\author{I. S. Kraev}
\affiliation{K.U.Leuven, Instituut voor Kern- en Stralingsfysica, Celestijnenlaan 200D, B-3001 Leuven, Belgium}%
\author{A. Knecht}
\affiliation{K.U.Leuven, Instituut voor Kern- en Stralingsfysica, Celestijnenlaan 200D, B-3001 Leuven, Belgium}%
\affiliation{PH Department, CERN, CH-1211 Geneva 23, Switzerland}
\author{T. Porobi\'{c}}
\affiliation{K.U.Leuven, Instituut voor Kern- en Stralingsfysica, Celestijnenlaan 200D, B-3001 Leuven, Belgium}%
\author{D. Z\'akouck\'y}
\affiliation{Nuclear Physics Institute, ASCR, 250 68 Re\v{z}, Czech Republic}%
\author{N. Severijns}
\affiliation{K.U.Leuven, Instituut voor Kern- en Stralingsfysica, Celestijnenlaan 200D, B-3001 Leuven, Belgium}%

%\address[kul]{Instituut voor Kern- en Stralingsfysica, KU Leuven, Celestijnenlaan 200D, B-3001 Leuven, Belgium}
%\address[cern]{PH Department, CERN, CH-1211 Geneva 23, Switzerland}
%\address[rez]{Nuclear Physics Institute, ASCR, 250 68 Re\v{z}, Czech Republic}

\begin{abstract}
Geant4 simulations play a crucial role in the analysis and interpretation of experiments providing low energy precision tests of the Standard Model. This paper focuses on the accuracy of the description of the electron processes in the energy range between 100 and 1000 keV. The effect of the different simulation parameters and multiple scattering models on the backscattering coefficients is investigated. Simulations of the response of HPGe and passivated implanted planar Si detectors to $\beta$ particles are compared to experimental results. An overall good agreement is found between Geant4 simulations and experimental data.
\end{abstract}

\maketitle

\section{\label{sec:intro}Introduction}

The search for physics beyond the Standard Model takes many forms. At the high energy frontier accelerators such as the LHC are able to produce new particles which could point toward new physics. The other, precision frontier relies on measurements of different observables, as e.g.\ in neutron and nuclear $\beta$ decay, where a deviation from the Standard Model value is an unambiguous and model independent sign of new physics \cite{Severijns2006,Abele2008,Nico2009,Konrad2010,Severijns2011,Severijns2013}. In order to further increase the precision of such measurements all possible systematic effects need to be evaluated, which often include Monte Carlo simulations such as the Geant4 simulation toolkit \cite{Agostinelli2003}. Among others it is widely used in neutron and nuclear correlation measurements \cite{Wauters2010,Rodriguez2009,Plaster2012,Li2013} and in searches for neutrinoless double-$\beta$ decay \cite{Bauer2006,Boswell2011}. 

The majority of these experiments are focusing on tracking and detection of electrons with typical $\beta$ decay energies (100~keV - 1~MeV) where one of the dominant systematic effects is the electron scattering from energy sensitive detectors. With the relative precision of the Standard Model tests in neutron and nuclear $\beta$ decay reaching the sub-percent level the accuracy of the Geant4 models needs to be re-evaluated and compared to new, high precision and high quality experimental data. 

This work focuses on the influence of the various Geant4 models and their parameters on the simulated values of the backscattering coefficients. It also investigates the quality with which experimental spectra of different $\beta$ decaying isotopes are reproduced. The results can be used to assign systematic errors to the simulations and also to estimate the systematic difference between simulated and experimental spectra.

\section{\label{sec:geant}Relevant Geant4 processes}
Geant4 \cite{Agostinelli2003} is a toolkit for simulating the passage of particles through matter. It was developed with the experiments at the LHC accelerator in mind and is therefore tuned to simulate high energy physics experiments. However, low-energy weak interaction experiments in neutron and nuclear $\beta$ decay that are dealing with $\beta$ particles of around 1~MeV kinetic energy typically, can also benefit from information provided by Geant4. At these low energies the physical processes involved are greatly reduced in number: practically only the electromagnetic interaction remains active. In this paper we will therefore focus on the electromagnetic processes of Geant4 \cite{Apostolakis2010, Amako2005}. Furthermore, we will focus on processes related to electrons. The relevant processes for this energy range used in Geant4 are the photoelectric effect, Compton-scattering and pair creation for $\gamma$ rays, while for electrons ionization, bremsstrahlung and scattering processes are included. Naturally all these processes are described by models, based on our current understanding of nature, but for practical purposes there will always be a compromise between realistic calculation time and the desired accuracy. A set of these models is called a physics list, and since version 9.3 of the Geant4 code these come in three flavors:
\begin{enumerate}
\item Standard - used for high energy physics experiments \cite{Burkhardt2004, Apostolakis2006}, but applicable to energies from 1~keV to 10~TeV \cite{Apostolakis2010}.
\item Livermore - extends the validity of the electromagnetic processes down to 250~eV, with more accurate descriptions of atomic effects and direct use of cross section data (the Standard physics list uses a parameterisation of these). This package used to be called the Low Energy physics list in version 9.0 and earlier \cite{Chauvie2004}.
\item Penelope - being developed based on the Penelope simulation package \cite{Sempau1997}, applicable to energies down to a few hundred eV \cite{geant4_phys_ref}. Note that this package does not include the Penelope-specific electron multiple scattering algorithm.
\end{enumerate}
It is to be noted that both Livermore and Penelope physics lists provide their own versions of several processes, such as ionization or bremsstrahlung \cite{geant4_phys_ref, geant4_loweprocesses_web}. However, the multiple scattering processes are shared with the Standard physics list. 

The multiple Rutherford scattering of electrons in matter is described by multiple scattering theories developed by Goudsmit and Saunderson \cite{Goudsmit1940} and later Lewis \cite{Lewis1950}. Both of these theories describe the individual scattering events by using Legendre polynomials, with their additive properties leading to an analytical solution of the final deflection angle after several scattering events. The Lewis theory provides the moments of the spatial displacement distribution as well.

Simulation of the individual Rutherford scattering events in Geant4 is possible by registering the Single Scattering \cite{Apostolakis2008, Ivanchenko2010} process to electrons and positrons. However, this is only practical for situations where the number of electron-electron collisions is low, e.g.\ for thin foils or low energy electrons. Since these conditions are in general not fulfilled, the multiple scattering (MSC) models were also implemented \cite{Kadri2009, geant4_phys_ref}. These models average out the individual scattering events thus allowing the steps to be longer and the simulation to run faster. The MSC models should therefore provide information about the angular deflection, true path length correction and spatial displacement of the electron. These models are not exact and are responsible for most of the electron transport uncertainties \cite{geant4_phys_ref}, affecting quantities such as the backscattering coefficient. 

The recent MSC models implemented in Geant4 are specific to a particle type, i.e. electrons, hadrons and muons. In this paper we will focus on the electron MSC models available in Geant4 version 9.5:
\begin{itemize}
 \item Single Scattering - simulates individual Rutherford scattering events, based on screened nuclear potentials, according to the Penelope code \cite{Fernandez-Varea1993}. 
 \item Urban MSC model - the default model within Geant4 is based on the Lewis MSC theory \cite{Lewis1950}, and is applicable to all particles. It uses model functions chosen such that they yield the same angular and spatial distributions as the Lewis theory. Furthermore, it uses different parameterisations of the central and tail part of the scattering angle distribution \cite{Ivanchenko2010}. Its performance has been validated against data obtained in thin foil transmission experiments \cite{Urban2006}. However, the majority of these experiments were carried out using primary beams of energies above 1~MeV.
 \item Goudsmit-Saunderson MSC model - based on the theory developed by Goudsmit and Saunderson \cite{Goudsmit1940} and is applicable only to electrons and positrons. It uses a database of cross sections generated by the ELSEPA code \cite{Salvat2005} and a sampling algorithm similar to the one presented in Ref.~\cite{Kawrakow1998}.
\end{itemize}
It is to be noted that the choice of the MSC model is independent of the physics list.

When simulating low energy experiments as e.g.\ $\beta$ decay, one first has to verify that the values of the various simulation parameters are suited for this energy range. The relevant parameters are \cite{Kadri2007,Elles2008,Ivanchenko2010}:
\begin{itemize}
\item Cut for Secondaries (CFS) - controls the way secondary particles are created, i.e. if a secondary particle would traverse in a given material a distance less than the CFS, it is not created but its energy is deposited locally. Therefore the value of this parameter should be smaller than the linear dimensions of the smallest ``sensitive'' geometrical volume. Its default value is 1~mm, however, with typical detector thicknesses of about 1~mm as used in low energy experiments this value is obviously too large and a CFS value of e.g.\ 1~$\mu$m is more suited, as was observed before \cite{Wauters2009b};
\item $F_R$ - limits the length of steps to a fraction of the electron mean free path. The default value is 0.04;
\item $F_G$ - determines the minimum number of steps in a given volume. The default value is 2.5;
\item Skin - dimensionless factor which defines a region near volume boundaries where single Coulomb scattering is applied. The thickness of this region is given by $\lambda \cdot$Skin, where $\lambda$ is the electron mean free path. The default value of this parameter is 3.
\end{itemize}

First, the performance of Geant4 with respect to electron backscattering will be investigated, considering the different physics lists, multiple scattering models and simulation parameters. Next, simulated spectra for different isotopes will be compared to experimental data obtained with both planar high purity germanium (HPGe) detectors and passivated implanted planar silicon (PIPS) detectors.

\section{\label{sec:backscatter}Backscatter comparisons}
\subsection{Introduction and literature review}
In order to validate the Geant4 electron processes one needs simple experiments (both in terms of geometry and of the physics involved) and high-quality data. A rather simple experimental observable related to electron processes is the backscattering coefficient. When an electron backscatters from the detector it deposits only part of its energy and then escapes from the detector, thereby distorting the shape of the measured electron spectrum. By simulating such a rather straightforward experiment the obtained backscattering coefficients can be directly compared to values cited in the literature.

\begin{figure}
\includegraphics[width=0.5\textwidth]{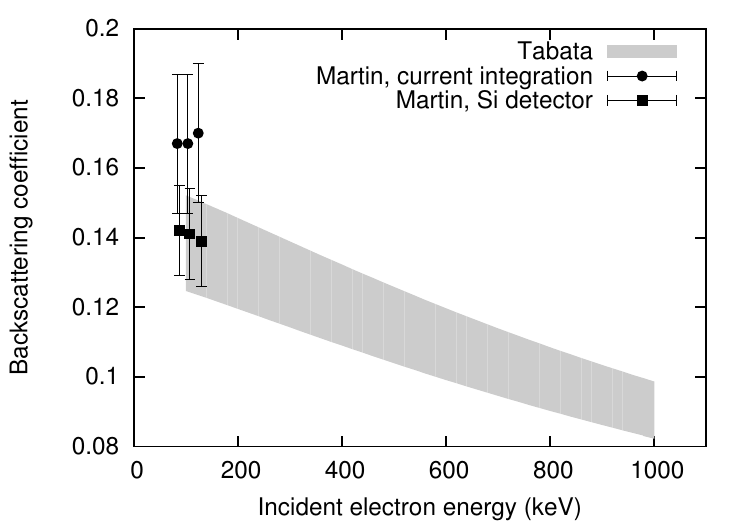}%
\caption{\label{fig:backscattertabata}Backscattering coefficients for normal incidence electrons on Si. The shaded region marked Tabata is the 1 $\sigma$ interval of the backscattering coefficient calculated using the formulas given by Tabata \cite{Tabata1971}. The data points are experimental results from Martin et al.\ \cite{Martin2003}, where the ``current integration'' and ``Si detector'' refer to the methods used to arrive to the results. }
\end{figure}

Tabata et al.\ \cite{Tabata1971} gave an empirical formula based on available experimental data (see Ref.~\cite{Tabata1971} and the references therein). The resulting backscattering coefficients for Si are shown as the shaded band in Figure~\ref{fig:backscattertabata}. Note that the uncertainties of the fitted parameters listed in Ref.~\cite{Tabata1971} induce relative uncertainties on the backscattering coefficients of about $10\%$ (width of the shaded band). Seltzer et al.~\cite{Seltzer1974} presented backscattering and transmission results for foils of various materials. The Geant4 MSC models were validated against these data. However, no data for Si or Ge were included in Ref.~\cite{Seltzer1974}. The most recent papers about backscattering of low energy electrons are by Martin et al.\ \cite{Martin2003, Martin2006}. These authors measured the electron backscattering coefficients for silicon, beryllium and organic scintillators in the energy range from 40~keV to 130~keV, and their results for Si are also shown on Figure~\ref{fig:backscattertabata}. It should be noted that no high-precision backscattering or transmission data for Si and Ge in the energy range between 150~keV and 1000~keV are currently available. However, the results from Martin et al.\ \cite{Martin2003, Martin2006} as well as our own previous work \cite{Wauters2010, Wauters2009b, Wauters2009,Wauters2009a} give good confidence in Geant4's ability to reproduce experimental data with an absolute precision that is typically of the order of 1~\%. 

\subsection{Simulations}
As a first step in investigating the performance of Geant4 with respect to electron backscattering we simulated a monoenergetic electron beam hitting a 1~mm thick slab of pure Si. Simulations were performed using Geant4 version 9.5 for all MSC models and the influence of the different simulation parameters listed in Section~\ref{sec:geant} were investigated as well. 
\subsubsection{Physics lists}
\begin{figure}
\includegraphics[width=0.5\textwidth]{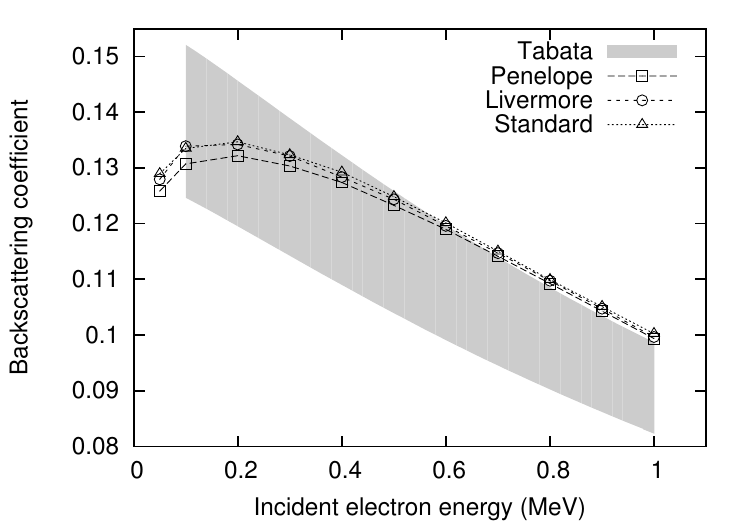}%
\caption{\label{fig:backscatt_physlist}Backscattering coefficient in function of the incoming electron energy. Data are shown for the Standard, Livermore and Penelope physics lists, using the default MSC model.}
\end{figure}

The backscattering coefficient as a function of energy obtained for the different physics lists is shown in Figure~\ref{fig:backscatt_physlist}. The differences are very small, with the Livermore and Standard physics lists providing almost identical results, demonstrating that for these simulations the usage of the low-energy packages (Livermore and Penelope) is not absolutely required. A notable feature of all three curves is the decrease of the backscattering coefficient below 200~keV, which is unrealistic. This is an artifact of the Urban MSC model, as will be demonstrated further on.
\subsubsection{Multiple scattering models}
\begin{figure}
\includegraphics[width=0.5\textwidth]{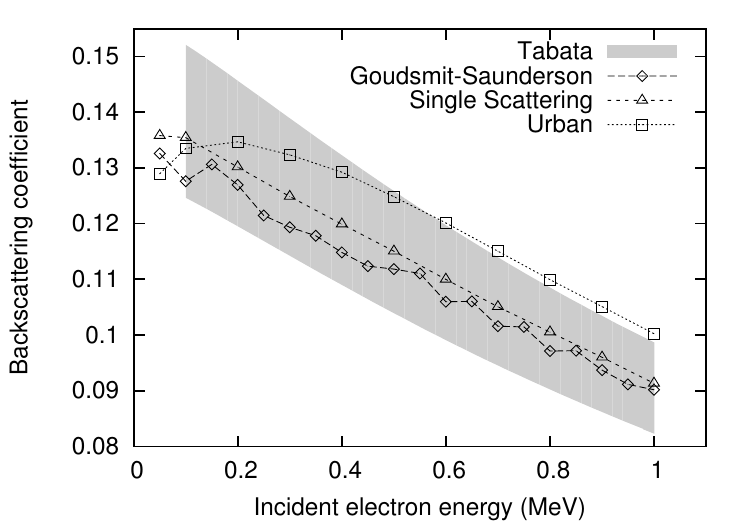}
\caption{\label{fig:backscatt_MSC}Backscattering coefficient as a function of the incoming electron energy, calculated with the Urban, Goudsmit-Saunderson and Single Scattering models within the Standard physics list.}
\end{figure}

As the backscattering results do not significantly depend on the physics list used, we will further use the Standard physics list to study the MSC models. As the backscattering coefficient depends on the accurate sampling of large scattering angles, we can expect larger differences between the condensed MSC models (Urban and Goudsmit-Saunderson) and the Single Scattering process. The results for all three models are shown in Figure~\ref{fig:backscatt_MSC}. For the Urban model a decrease in the backscattering coefficient is observed below 200~keV. This, in combination with other possibly unidentified effects, contributes to the fact that the difference between simulated and experimental spectra rises up to 50\% in the energy region below 100~keV (see Sections~\ref{sec:hpgedet} and \ref{sec:pipsdet}). Further, a constant offset is visible compared to the other models as well as to the empirical relation; the reason for this is unclear. The Single Scattering model shows more stable behavior and also yields values closer to the central values obtained from the empirical relation of Tabata et al.~\cite{Tabata1971}, although the 10\% uncertainty on the values calculated with this relation does not exclude the two other models. The Goudsmit-Saunderson model is found to exhibit a clear ``staggering'' effect which is not expected from physics grounds, rendering this model less interesting for applications that require high precision.

\subsubsection{Simulation parameters}
\begin{figure}
\includegraphics[width=0.5\textwidth]{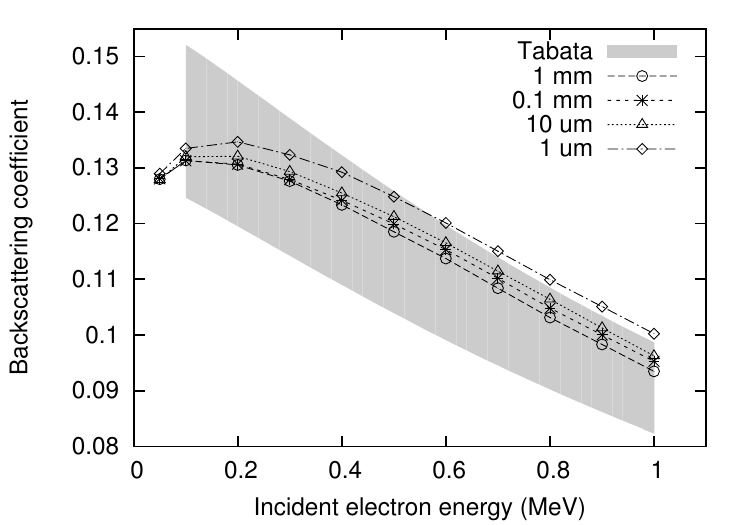}
\caption{\label{fig:backscatt_CFS}Backscattering coefficient as a function of the incoming electron energy for different values of the CFS parameter. Simulations were performed using the Standard physics list with the Urban MSC model.}
\end{figure}

\begin{figure}
\includegraphics[width=0.5\textwidth]{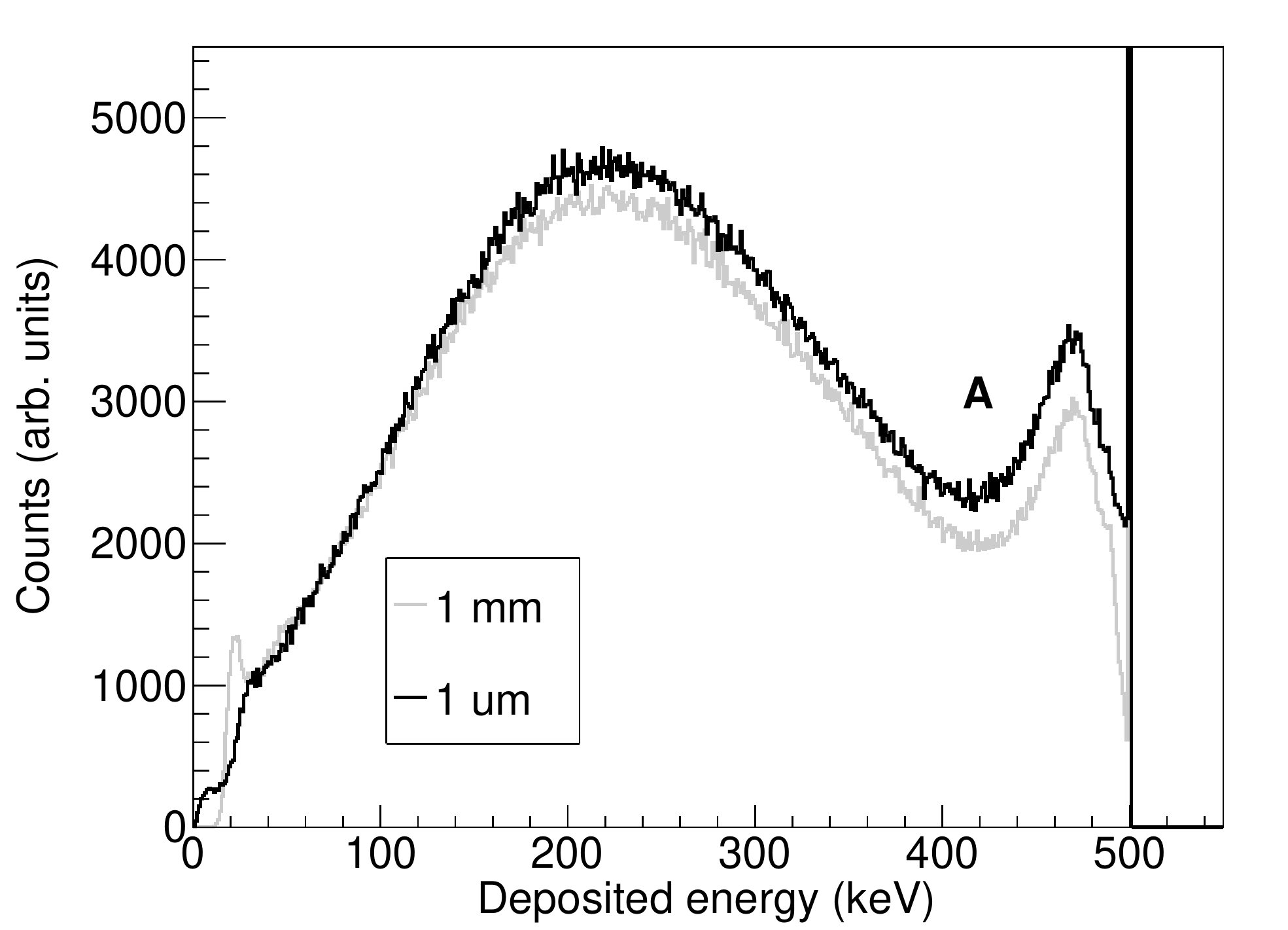}%
\caption{\label{fig:spectrum_CFS}
Spectrum of deposited energies for 500~keV incoming electrons. The default value for CFS of 1~mm is compared to our recommended value of 1~$\mu$m.}
\end{figure}

To investigate the effect of the different simulation parameters on the backscattering coefficient the Urban MSC model was used  as it is the default model and also requires the smallest calculation time. The dependence of the backscattering coefficients on the CFS value is shown in Figure~\ref{fig:backscatt_CFS}. The fact that the backscattering coefficient increases with decreasing CFS value is expected (the smaller the CFS the more low energy secondaries are created). However, for a value of 1 $\mu$m it reaches the edge of the 1 $\sigma$ band of Tabata. To further investigate this difference, spectra of deposited energies for the backscattered events are shown in Figure~\ref{fig:spectrum_CFS} for the extreme cases CFS~=~1~$\mu$m and CFS~=~1~mm. The effect of the different values of the CFS parameter is as expected, i.e. for smaller CFS values the probability for the electron to deposit a higher fraction of its initial energy increases (feature ``A'' on Figure~\ref{fig:spectrum_CFS}). Despite the slightly worse agreement between the Tabata values \cite{Tabata1971} and the simulated data for CFS~=~1~$\mu m$ we will for the time being continue to use the value of 1~$\mu$m for this parameter as it is considered more realistic for our purposes (e.g.\ the detector dead layer thicknesses are typically of the order of several 100~nm).


\begin{figure}
\includegraphics[width=0.5\textwidth]{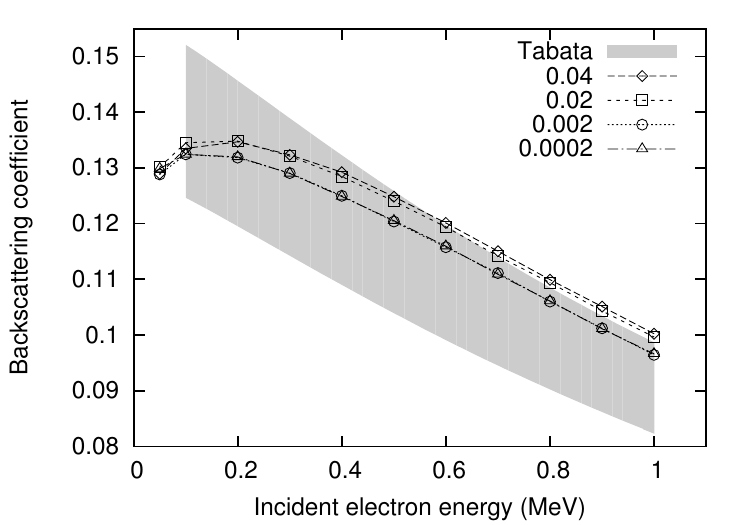}
\caption{\label{fig:backscatt_rf}Backscattering coefficient as a function of the incoming electron energy for different values of the $F_R$ parameter. Simulations were performed using the Standard physics list with the Urban MSC model.}
\end{figure}

\begin{figure}
\includegraphics[width=0.5\textwidth]{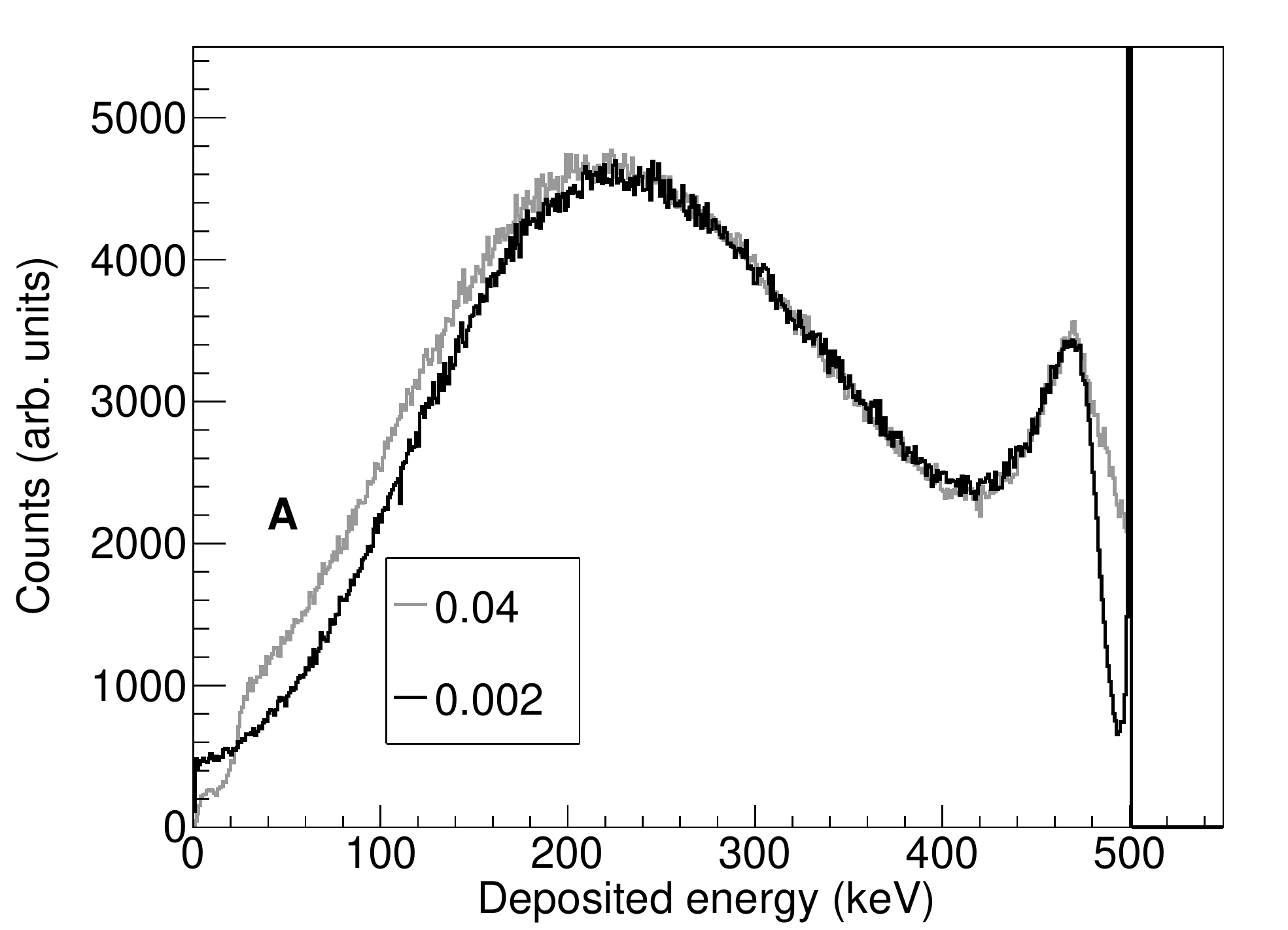}%
\caption{\label{fig:spectrum_rf}Spectrum of deposited energies for pure backscatter events for 500~keV normal incident electrons. The default value for $F_R$ of 0.04 is compared to the value of 0.002.}
\end{figure}

Simulation results for different values of the $F_R$ parameter are shown on Figure~\ref{fig:backscatt_rf}. The backscattering coefficient is found to saturate when the $F_R$ parameter drops below 0.002 (see Figure~\ref{fig:backscatt_rf}). The corresponding spectrum of deposited energies for the backscattered electrons (Figure~\ref{fig:spectrum_rf}) shows less events for incident electrons depositing only a small fraction of their initial energy (feature ``A'' in Figure~\ref{fig:spectrum_rf}) in the detector before being backscattered. Further, for $F_R$~=~0.04 a sharp drop is observed below about 30~keV, which is not physical. Therefore, an $F_R$ value between 0.01 and 0.002 seems realistic. A dedicated experiment focusing on these effects, so as to determine the best value, would be welcome.

Simulations performed for different values of the $F_G$ and Skin parameters did not result in significant changes of the backscattering coefficient. Therefore, in subsequent simulations their default values (see Section~\ref{sec:geant}) were used.

\subsection{Conclusions on the backscattering coefficients}
The different MSC models and simulation parameters influence the backscattering coefficient at the 10\% level, which is approximately equal to the uncertainty of the empirical relation of Tabata et al.\  \cite{Tabata1971}. The default value of the CFS parameter of 1~mm is too large if one uses typical particle detectors with thicknesses up to several mm. Furthermore, the typical dead layer thicknesses are much smaller, such that a more realistic value is around 1~$\mu$m. Simulations using this value for the CFS parameter produce backscattering coefficients in agreement with the results from the empirical relation. Based solely on the backscattering coefficients the best value for $F_R$ can not be determined unambiguously. We therefore fix it for the time being at the default value of 0.04. The Single Scattering model agrees the best with the empirical equation of Tabata, but unfortunately it requires approximately 10 times more computer time than the Urban or the Goudsmit-Saunderson models.

Since the MSC models need to provide the angular deflection (scattering angle) after each step, a suitable benchmark would be high precision data on the angular distribution of electrons after transmission through thin foils of various thicknesses \cite{Urban2012}. We are in the process of preparing such an experiment.

\section{\label{sec:detector}Simulation of detector response}
A Geant4 simulation records the energy deposited in a specified geometrical volume. Such a simulated spectrum, however, can not be directly compared to the measured spectrum since several instrumental effects as well as the decay scheme of the isotope considered still have to be taken into account.

\subsection{Energy resolution}
Geant4 does not take into account effects such as charge trapping in the detector or noise originating in the preamplifier and in the amplifier, all of which determine the energy resolution observed. Although Geant4 provides built-in classes to address these issues, we prefer another approach. The net effect of these random changes to the signal is best described by a Gaussian spread of the final simulated spectrum from Geant4. For the width of the Gaussian used to convolute that spectrum we use a value determined by a $\chi^2$ fit to the conversion electron peaks from the decay of $^{207}$Bi at 482 and 976~keV.

\subsection{Pile-up}
As will be seen in the following sections, even for a pure $\beta$ spectrum, i.e. with no $\gamma$-rays being present in the decay scheme, often events above the endpoint energy are observed (see e.g.\ Figure~\ref{fig:85kr_simexp}). This is due to detector event pile-up, an artifact of signal processing. This effect can be accounted for in several ways. We prefer to deal with this in post processing, since it is then easy to change the pile-up probability thus accounting for changes in the source activity as, e.g., occurs in on-line experiments. One then introduces a probability that two random events from the spectrum are summed together. The magnitude of this probability is determined by the best fit to the region of the experimental spectrum above the $\beta$ endpoint energy, or if a pulser peak is present, by the pulser peak-to-tail ratio.

\subsection{Geant4 Radioactive Decay}
Geant4 handles nuclear decays via the \emph{G4RadioactiveDecay} process. This includes the different decay modes with their branching ratios and automatically generates decay products, such as $\alpha$ or $\beta$ particles. Considering $\beta$-decay, besides the phase space factors only the Fermi function is implemented however. Therefore, in the studies concerning $\beta$ decaying isotopes a custom made code was used \cite{Wauters2009b} with the Fermi function and all higher order corrections implemented according to the prescriptions of Wilkinson \cite{Wilkinson1989,Wilkinson1990,Wilkinson1995,Wilkinson1997}. 

\section{\label{sec:hpgedet}Geant4 performance for planar HPGe detectors}
\begin{figure}

\includegraphics[width=0.5\textwidth]{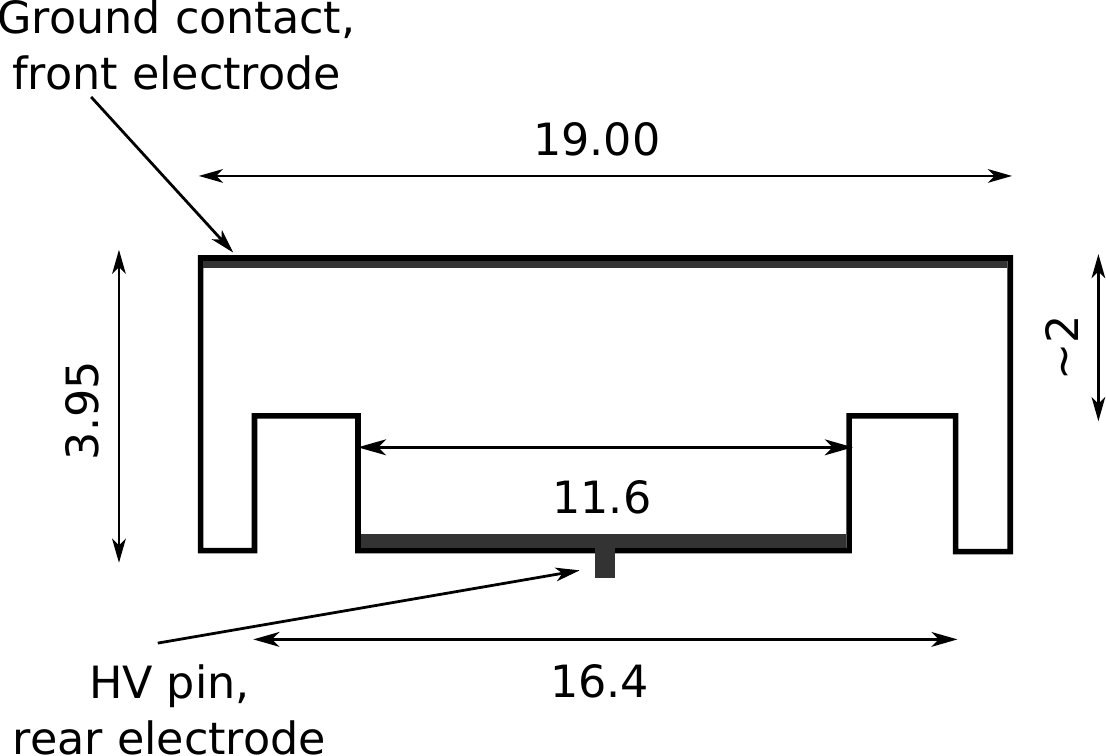}%
\caption{\label{fig:HPGeDimensions} Dimensions of the 15/4 HPGe particle detector. All numbers are in units of mm.}
\end{figure}

Custom made planar HPGe detectors were developed \cite{Venos1995,Venos2000} for low temperature nuclear orientation $\beta$-asymmetry measurements. They were used in an experiment with $^{114}$In \cite{Wauters2009} and in the $^{67/68}$Cu experiment at ISOLDE, CERN  \cite{Soti2013a}. All of them were extensively tested \cite{Wauters2009thesis}, and in this paper we will focus on the 15/4 detector. Figure~\ref{fig:HPGeDimensions} shows a sketch of the detector with all its dimensions noted. The detector has its front electrode made with boron implantation and the thickness of this dead layer is estimated to be $\sim 100$~nm. The thickness of the Li diffused dead layer at the rear electrode side of the detector was measured to be in the range of 0.7-0.9~mm. Simulations showed that a variation of 0.1~mm in this thickness does not change the response of the detector significantly. The sensitive area of the detector was modeled according to the results of a series of measurements with collimators of different size, further supported by COMSOL-Multiphysics \cite{Comsol} simulations.

The detector was positioned inside a vacuum chamber and a 1~mm thick Cu collimator with a 12~mm diameter circular hole was mounted in front of it. The role of this collimator is to stop the electrons arriving at the edge of the detector where the electric field might not be uniform and thus not all the charge created would be collected. The response of the detector was extensively tested at a temperature of 77~K with four different radioactive sources, i.e. $^{60}$Co, $^{85}$Kr, $^{90}$Y and $^{207}$Bi. The $^{207}$Bi source was sandwiched between two 5.325~$\mu$m thick Ti foils in which the conversion electrons loose only about 2~keV energy. The $^{60}$Co source was sandwiched between two, 10~$\mu$m thick mylar foils, leading to an energy loss for electrons of about 3-4~keV. The $^{85}$Kr source was a 0.05~mm thick iron foil in which the radioactive nuclei were implanted up to a depth of around 15~nm. The $^{90}$Y source was prepared in-house by drying a small drop of liquid solution containing $^{90}$Sr (which decays to $^{90}$Y) inside a hole in a piece of aluminum of $25 \times 10 \times 1 \textrm{mm}^3$. Thereafter the activity was covered with a 0.1~mm thin layer of epoxy.

\subsection{\label{sec:hpge_comp} Comparisons with Geant4}
\begin{figure}
\includegraphics[width=0.5\textwidth]{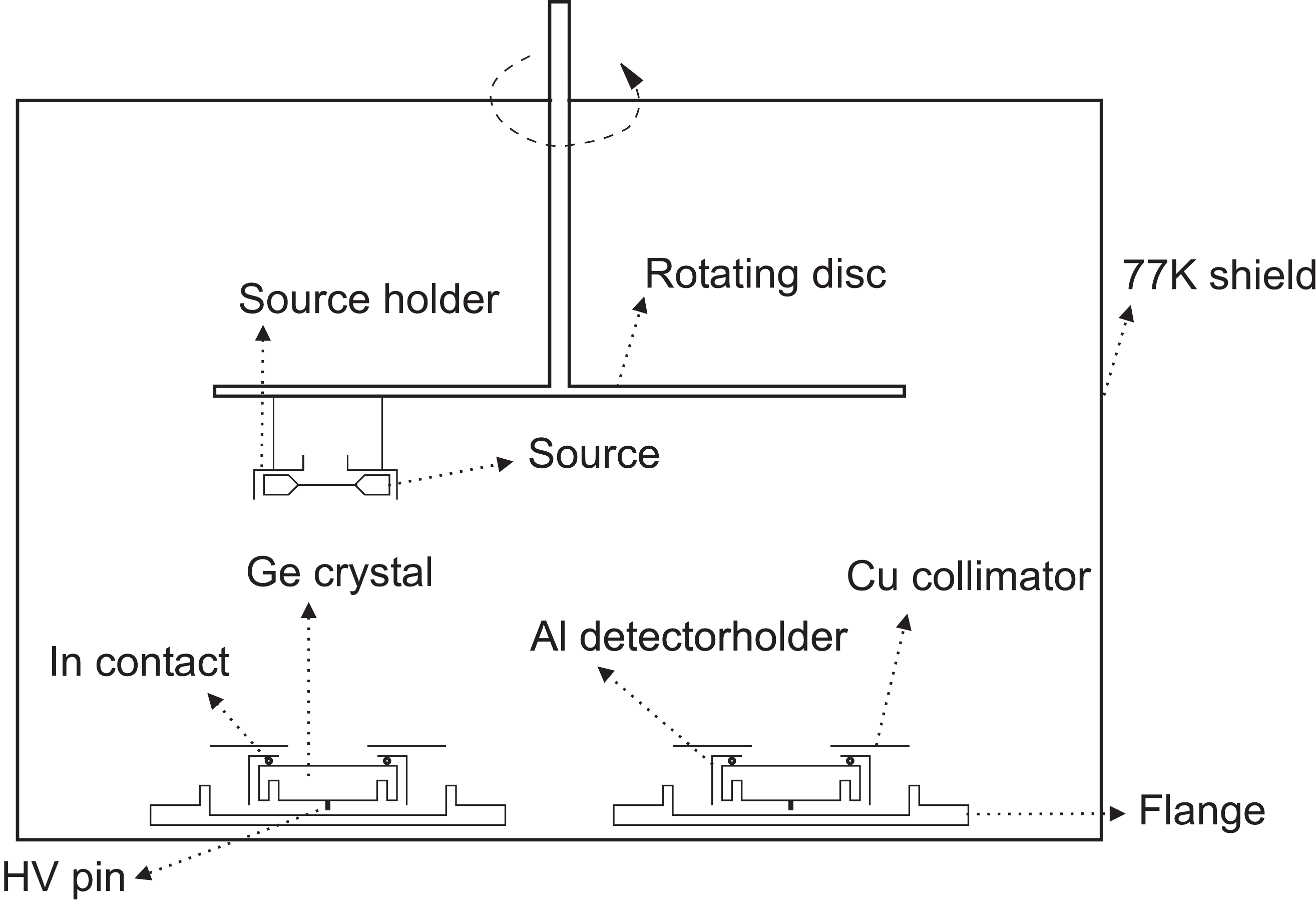}
\caption{\label{fig:HPGe_setup}
Sketch of the experimental setup used to measure electron spectra of different isotopes with HPGe detectors. The bottom plate of the vacuum chamber is connected to a liquid nitrogen bath allowing the detectors to be cooled to 77~K. The radioactive source is mounted on a rotating plate so it can be positioned above any of the detectors without the need for opening the system.}
\end{figure}

The detailed geometry of the entire setup (vacuum chamber, support structures,  sources and detectors, see Figure~\ref{fig:HPGe_setup}) used to measure electron spectra with the different sources was implemented in Geant4. Simulations were then performed for each detector-source combination and the resulting histograms were normalized to the number of counts in an energy range depending on the isotope. For comparing the experimental to the simulated spectra we used the reduced $\chi^2$ defined as:
\begin{equation}
\label{eq:chidef}
\chi^2_{red} = \frac{1}{\nu}\sum\limits_{i}{\frac{(y_i^{exp} -y_i^{sim})^2}{\sigma_{i,exp}^2+\sigma_{i,sim}^2}}
\end{equation}
with $\nu$ the number of degrees of freedom, $y_i^{exp}$ and $y_i^{sim}$ the content of the $i$th bin in the experimental and simulated spectrum, respectively, and $\sigma$ the associated uncertainty. In the ideal case $\chi^2_{red}$ should be equal to unity. However, the measurements presented in this paper were performed to investigate the impact of the different Geant4 parameters, so that differences between experimental data and simulations larger than the statistical uncertainties can be expected. Although the $\chi^2_{red}$ value can thus not be expected to be around unity, it can, however, still be considered as a relative figure of merit between  simulations for different parameters or models being used.

\subsubsection{$^{207}$Bi}
\begin{figure}
\includegraphics[width=0.5\textwidth]{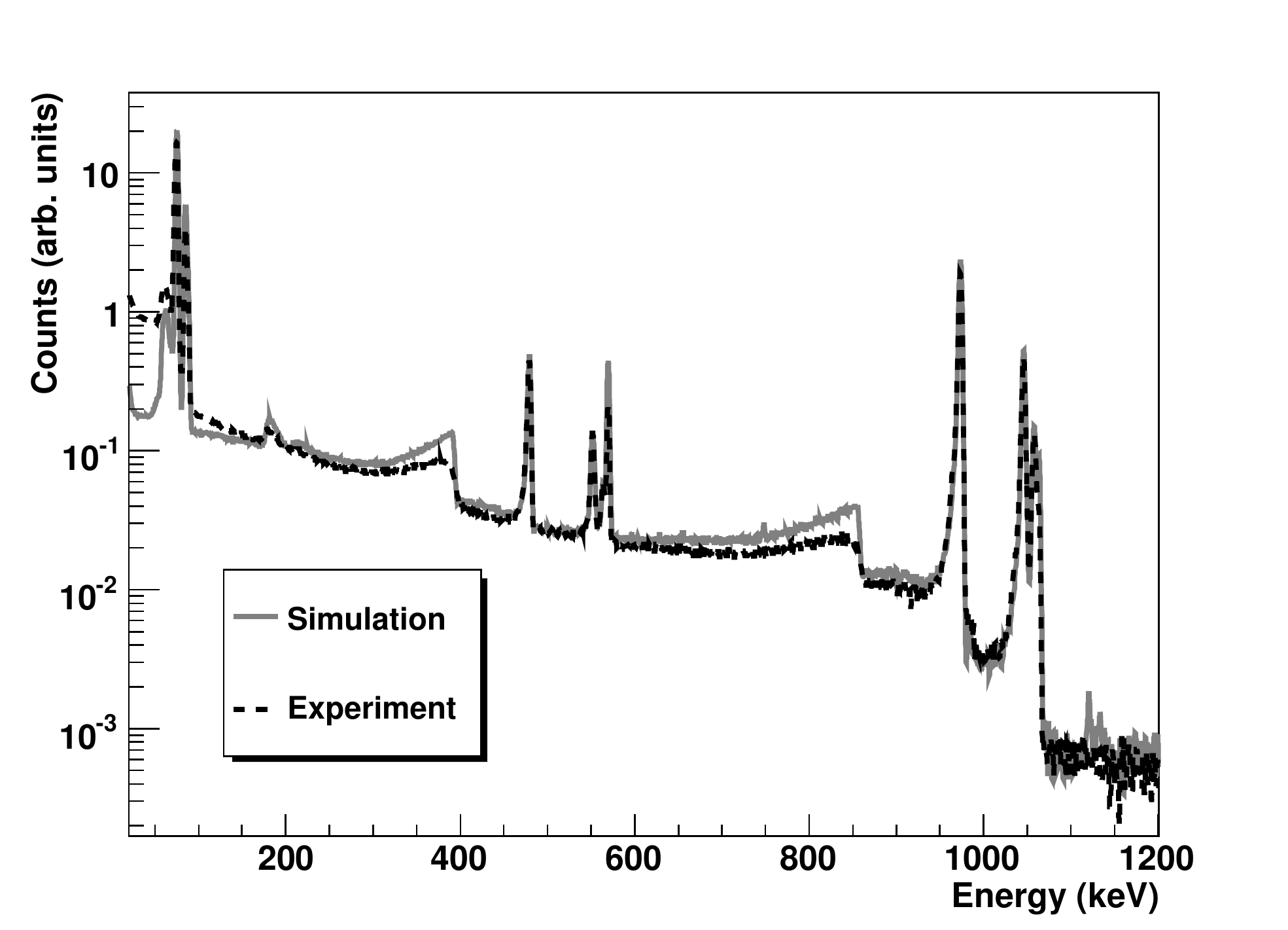}
\caption{\label{fig:bi_sim_exp} Comparison between experimental and simulated spectra of $^{207}$Bi, for the 15/4 detector. The two main $\gamma$ lines (at 569.7~keV and 1063~keV) are visible (highest energetic peak of the groups around 500~keV and 1000~keV), together with their K, L (and M) conversion electrons. Both $\gamma$ rays generate Compton edges which are located at 394 and 857 keV, respectively. The K$\alpha$ and K$\beta$ X-ray lines are at 74~keV and at 85~keV. The spectra were normalized in the energy region from 50 to 1200~keV.}
\end{figure}

Reproducing the experimental spectrum obtained for this isotope is the most demanding job for Geant4 as the decay of $^{207}$Bi produces X-rays, conversion electrons and $\gamma$ rays over a wide energy range. Spectra obtained with these thin detectors are dominated by the conversion electrons, while the $\gamma$ rays contribute mainly via the Compton effect. As this isotope decays via electron capture and because of its relatively complex decay scheme the standard radioactive decay module of Geant4 was used in the simulations. In Figure~\ref{fig:bi_sim_exp} the experimental and simulated spectra for the 15/4 detector are compared. Although the overall features are well reproduced for energies above 150~keV, clear differences between simulation and experiment are observed in some parts, especially near the Compton edges which are overestimated by the simulation. 

The distinct difference between simulation and experiment in the spectrum of $^{207}$Bi near the Compton edge of the two $\gamma$ rays might be in part due to the inaccuracy of the Compton scattering cross sections, which is further emphasized by the relatively high $Z$ value of Ge. As the $\gamma$ processes within Geant4 have been validated at the level of several percent \cite{Cirrone2010} the reason for the observed difference is most probably an interplay between several effects, such as the e.g.\ the fine details of the Compton scattering process and the detection of the resulting electron. Note that no clear dependence of the Compton-edge intensity on the detector thickness was observed. It is to be noted that this effect was also observed when comparing spectra measured with PIPS detectors (see Section~\ref{sec:pipsdet} and Figure~\ref{fig:pips_bi_ComptonEdge}), although less pronounced.

It was found that simulations performed with the Penelope or Livermore physics lists produced a smoother Compton edge, similar to Figure~\ref{fig:pips_bi_ComptonEdge}. However, the differences between the Standard and Penelope (or Livermore) physics lists is significantly smaller than the difference between simulation and experiment.

\subsubsection{$^{85}$Kr}
\begin{figure}
\includegraphics[width=0.5\textwidth]{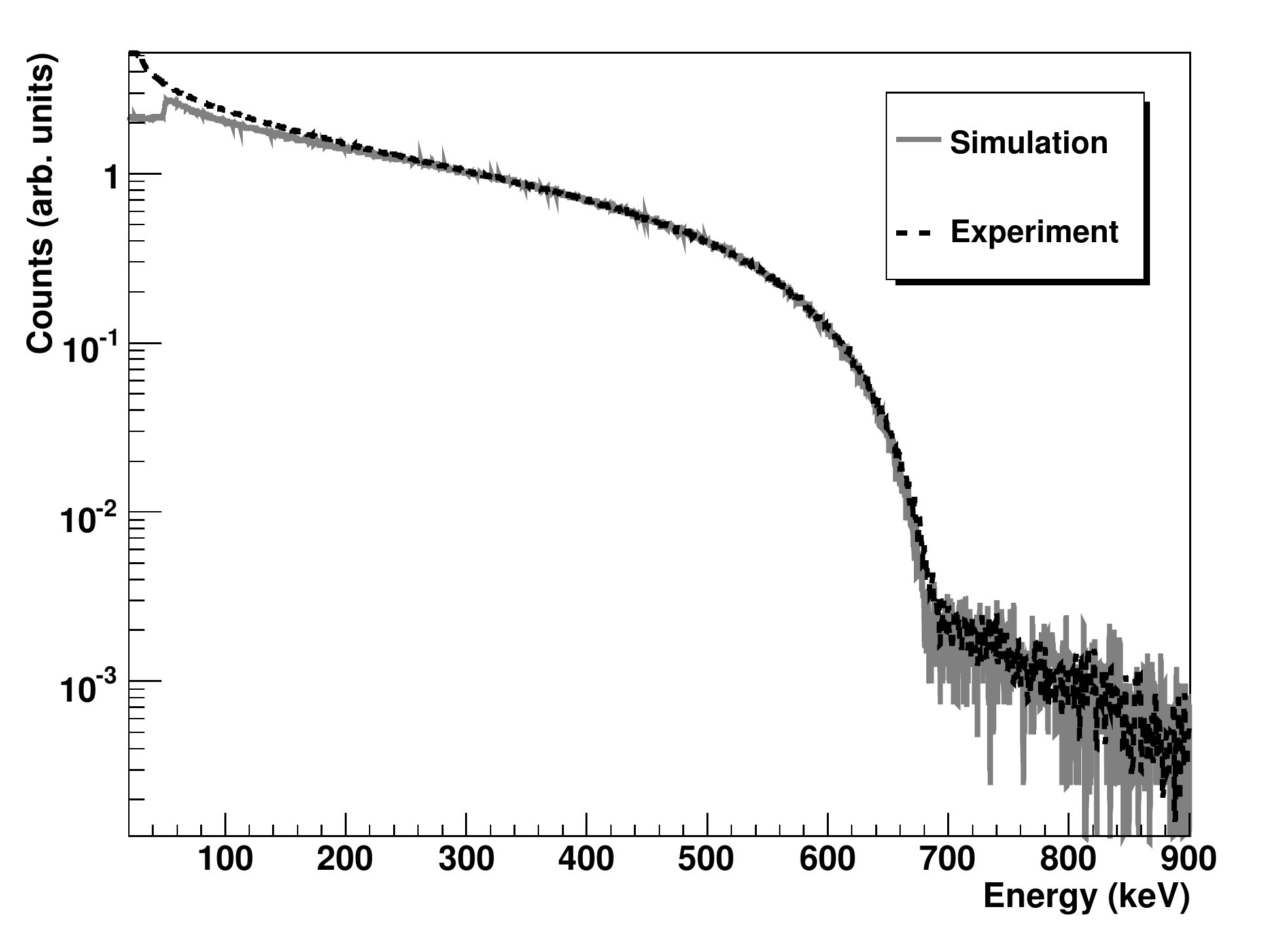}
\caption{\label{fig:85kr_simexp}
Comparison between the simulated and measured $^{85}$Kr spectra for the 15/4 detector.
The normalization region is 300-600~keV.
In this region $\chi^2_{red} = 1.4$ with 74 degrees of freedom and simulation and experiment agree within $2\%$.
}
\end{figure}

This isotope is suitable to check detector response to relatively low energy $\beta$ particles (the endpoint energy of the $^{85}$Kr decay is $E_0 = 687$~keV), without any disturbance from $\gamma$ rays. However, the radioactive decay module of Geant4 does not generate the correct spectrum shape for this isotope as it does not decay via an allowed $\beta$ transition but via a so-called first forbidden unique $\beta$ transition, the spectrum shape of which is not included in Geant4. For this type of transitions the spectrum shape differs from the allowed one by a factor \cite{Langer1949} 
\begin{equation}
 (W^2-1) + (W_0-W)^2
\end{equation}
with $W$ the total energy of the $\beta$ particle and $W_0$ the total endpoint energy, both in units of the electron rest mass $m_ec^2$. After implementing the necessary correction factors the experimental and simulated spectra are found to agree within 2\% (see Figure~\ref{fig:85kr_simexp}).

\subsubsection{$^{90}$Y}
\begin{figure}
\includegraphics[width=0.5\textwidth]{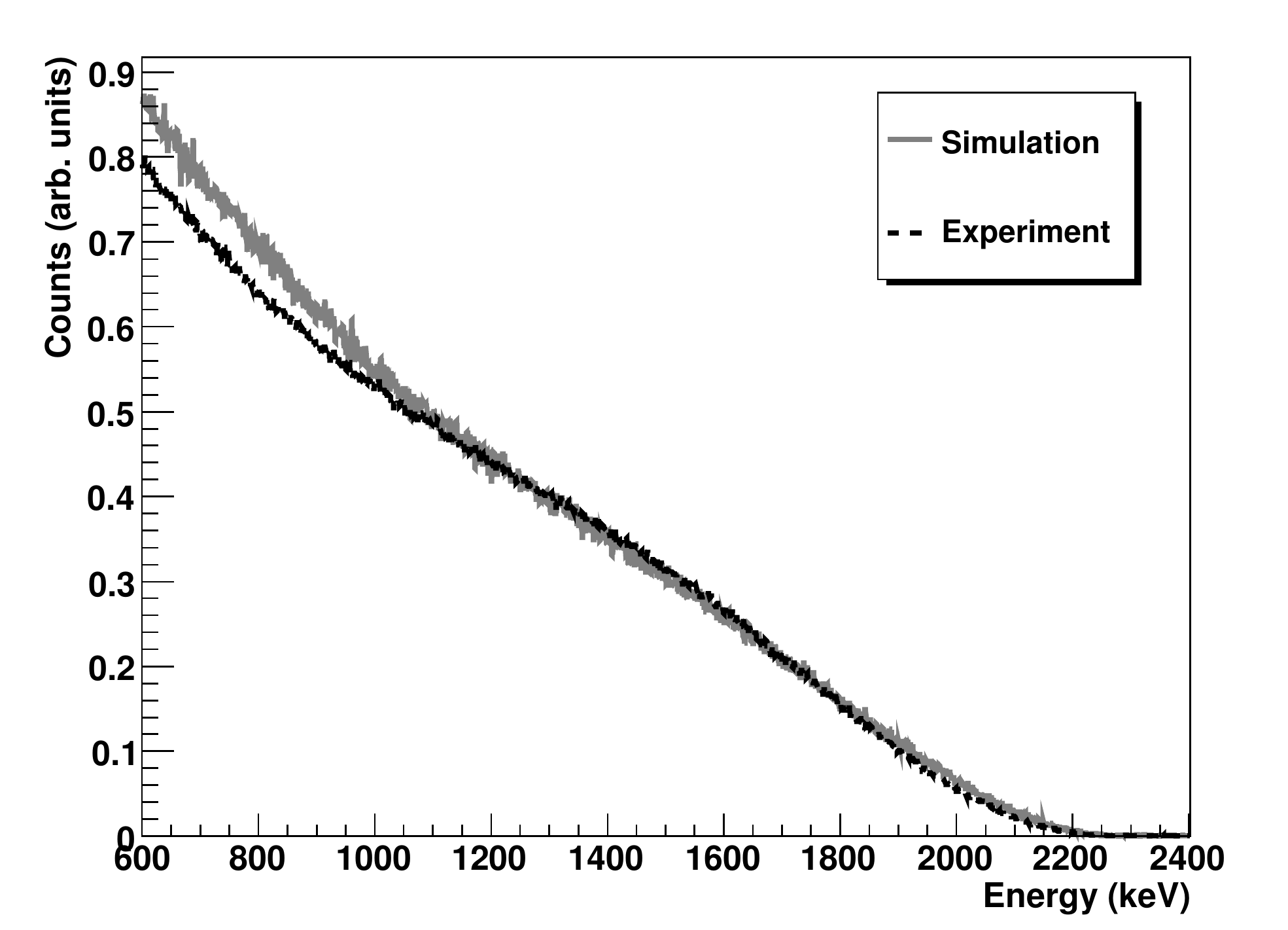}
\caption{\label{fig:90Sr_simexp}
Comparison between the simulated and measured $^{90}$Sr spectra for the 15/4 detector. 
The normalization region is 1200-2000~keV. In this region $\chi^2_{red} = 2.3$ with 532 degrees of freedom. }
\end{figure}

The isotope $^{90}$Y ($\beta$ endpoint energy $E_0$~=~2.2~MeV) was obtained as the decay product of $^{90}$Sr ($\beta$ endpoint energy $E_0$~=~546~keV). Only the decay of $^{90}$Y was simulated and simulation and experiment were only compared in the part of the $^{90}$Y $\beta$ spectrum above the $^{90}$Sr $\beta$ endpoint energy. As $^{90}$Y  decays via a first forbidden unique $\beta$ transition the same correction factors were applied to the Geant4 spectrum generator as in the case of $^{85}$Kr. The accuracy of the simulations is rather limited for energies below 1~MeV (Figure~\ref{fig:90Sr_simexp}). The upper 1~MeV of the spectrum can be reproduced with $\sim 5\%$ precision. Besides the fact that this is a first forbidden unique transition this observed difference could be in part due to the fact that the exact geometry of the source was known with much less precision compared to the other sources. Previous measurements \cite{Wauters2009b} with $^{60}$Co have shown that the measured spectrum is indeed rather sensitive to the detailed geometry of the source. 

\subsubsection{\label{sec:hpge_60co}$^{60}$Co}
This isotope is very well suited to test the performance of the Geant4 code for low energy $\beta$ particles due to its relatively simple decay scheme. The $\beta$ endpoint energy is 318~keV which is much lower than the energies of the two strong $\gamma$ lines (1.173 and 1.332~MeV) in the decay of this isotope. The problem of subtracting the Compton background caused by these $\gamma$ rays significantly contributed to the error budget in past experiments \cite{Wauters2010}.

\begin{figure}
\includegraphics[width=0.5\textwidth]{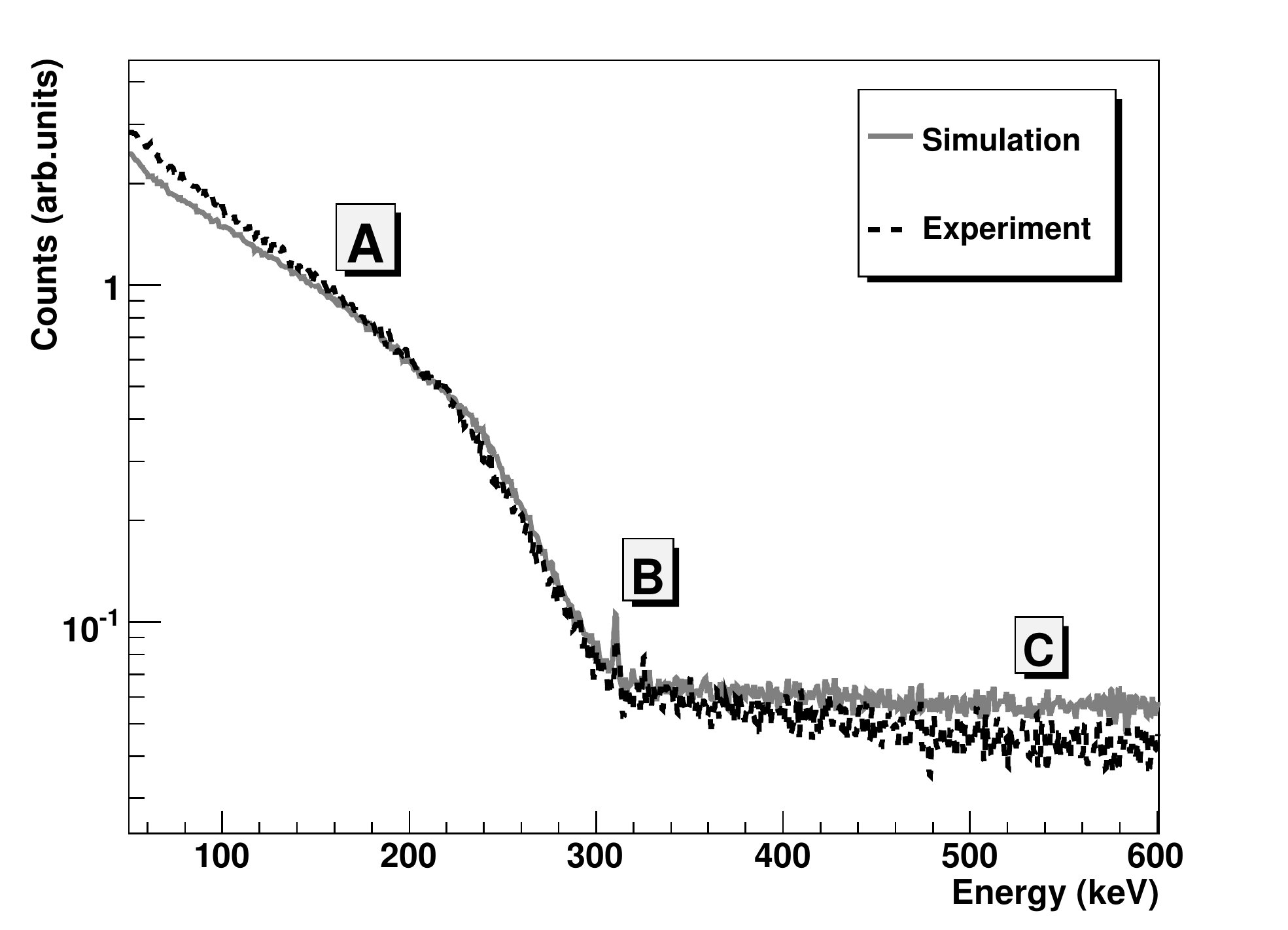}
\caption{\label{fig:60Co_simexp}
Comparison between the simulated and measured $^{60}$Co spectra for the 15/4 detector.
The small peak (marked ``B'') at 310~keV is the double escape peak of the 1332~keV $\gamma$ line. 
The normalization region is 150-300~keV.
In this region $\chi^2_{red} = 2.8$ with 150 degrees of freedom. }
\end{figure}

Figure~\ref{fig:60Co_simexp} shows the upper part of the beta spectrum together with part of the Compton background. In the ``A'' region (from 150~keV to 300~keV) the difference between simulation and experiment is around $5\%$. At energies below 150~keV the difference becomes much larger. This can be due to several reasons:
\begin{itemize}
 \item a problem with simulating the backscattering of low energy electrons (see sec.~\ref{sec:backscatter}), or
 \item the Compton plateau not well being reproduced by the simulations (see sec.~\ref{sec:60Co_pips}).
\end{itemize}
A significant difference between simulation and experiment can be observed in the intensity of the Compton background (region ``C''), where the simulation shows a clear excess of counts. This effect was observed before with Si detectors \cite{Wauters2010,Wauters2009b} as well.

\subsection{Conclusions for HPGe detectors}
The general features of all measured spectra are rather well reproduced by the Geant4 simulations. The high energy part of the $\beta$ spectra are typically reproduced at the \mbox{2-3\%} level. However, the lower half of the $\beta$ spectra are typically much less well reproduced in simulations. A possible reason for this could be the inaccuracy of the backscattering coefficients at these energies, see Section~\ref{sec:backscatter}. 

Past evaluations of Geant4 performance in simulating spectra of HPGe detectors were mostly carried out for large volume detectors mostly employed for $\gamma$ detection \cite{Boswell2011,Schumaker2007,Hurtado2004}. The observed accuracy of several percents on the $\gamma$-peak efficiencies is slightly better than what is found here. 

The observed differences between the experimental and simulated spectra for a HPGe detector are found to be much larger than the small effects related to the choice of physics list, MSC models or of values for the Geant4 parameters. Therefore, unfortunately, no additional information on the best values of these parameters can be obtained in this case. 

\section{\label{sec:pipsdet}Geant4 performance for PIPS detectors}
A 1.5~mm thick, fully depleted pure PIPS detector (MSX03-1500, from Micron Semiconductor) was recently tested by our group, in part to replace the Hamamatsu 0.5~mm thick PIN diode detectors \cite{Wauters2010, Wauters2009}. The front dead layer of this detector is 100~nm thick. The entrance window consists of a 300~nm thick Al grid with a 3\% coverage. This detector is well suited for precision $\beta$ spectroscopy because the low $Z$ values of Al and Si limit the probability for electron backscattering from the entrance window. It is further able to fully stop up to 800~keV electrons.

\begin{figure}
\centering
\includegraphics[height=0.3\textwidth]{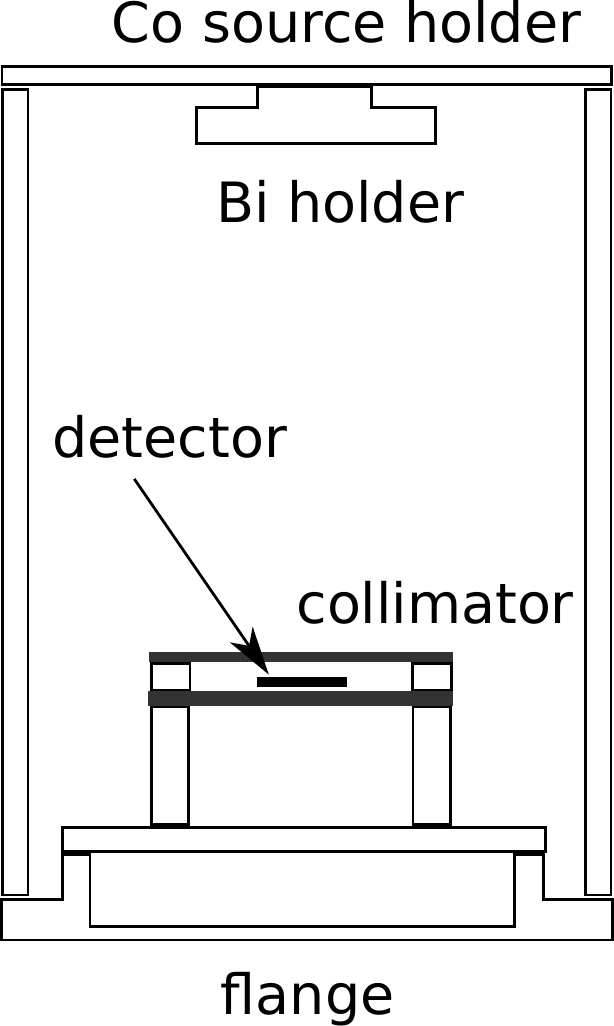}
\caption{\label{fig:pips_setup} Sketch of the experimental setup used to measure $^{207}$Bi and $^{60}$Co spectra with the PIPS detector. The detector-source distance is 71~mm. The entire setup was positioned in a cryostat and cooled to 77~K in order to reach full detector depletion. }
\end{figure}

The detector was tested in the same vacuum chamber used for testing the HPGe detectors and again at 77~K (see Section~\ref{sec:hpgedet} for more details). A 0.8~mm thick Cu collimator with a 9~mm diameter circular hole was mounted in front of the detector. 

\begin{figure}
\includegraphics[width=0.5\textwidth]{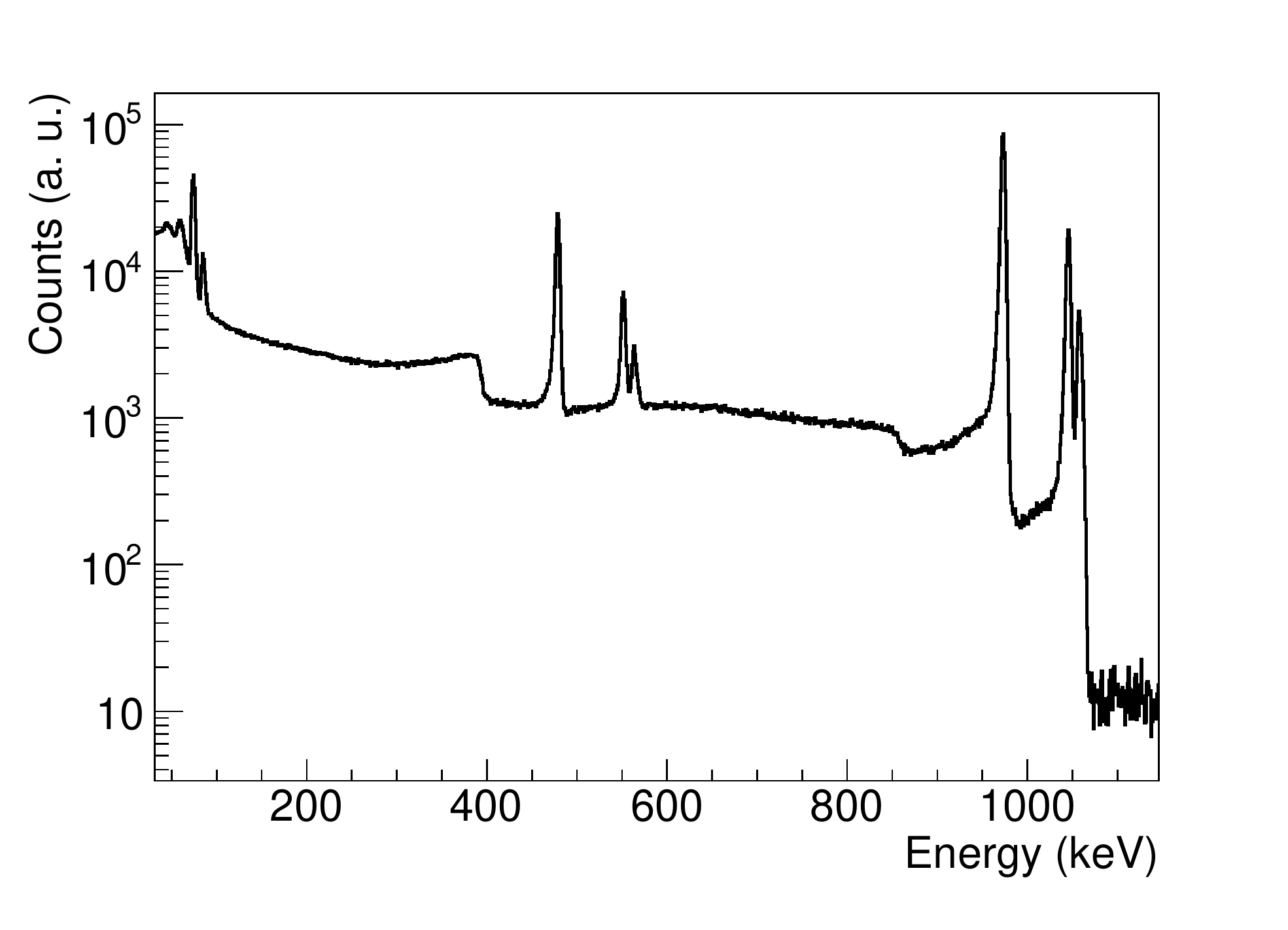}%
\caption{\label{fig:pips_bi_exp} The $^{207}$Bi spectrum registered with the PIPS detector operating at 77~K. The two main $\gamma$ lines (at 569.7~keV and 1063~keV) are not visible, but their K, L and M conversion electrons generate the two groups of peaks. The Compton edges of both $\gamma$ rays are also visible. The K$\alpha$ and K$\beta$ X-ray lines are visible at 74~keV and at 85~keV, respectively. Detector resolution (FWHM) is 4~keV at 482~keV.}
\end{figure}

\subsection{Depletion}
The spectra of $^{207}$Bi measured with the detector at room temperature showed signs that the detector was not fully depleted. The conversion electron peaks were not of the expected intensity. Indeed, the experimental ratio of the K conversion electron lines K$_{1063}$/K$_{569}$ was 0.998(14), in clear disagreement with the expected value of about 3.8, when assuming full depletion and taking into account the detection efficiency and backscattering probability at the different energies. The results of Geant4 simulations were supporting the assumption of partial depletion at room temperature, allowing to estimate the depletion thickness to be around 0.8 to 0.9~mm. After cooling the setup to 77~K (Figure~\ref{fig:pips_bi_exp}) the experimental conversion peak ratio became 3.78(11), fully consistent with the expected value and with the error being dominated by the uncertainties of the conversion coefficients \cite{Kondev2011}. In order to confirm full depletion, simulations were performed for slightly different depletion layer thicknesses. The comparison of these with experiment is summarized in Table~\ref{tab:pips_depl}. Values for the K conversion peak ratio very close to the experimental one are found. The best $\chi^2_{red}$ for the comparison of simulated and experimental spectra is obtained for depletion thicknesses of about 1.50~mm, thus confirming the probably full depletion. Note that it was observed that this detector can also be operated with no significant drop in performance at liquid helium temperature (4~K) making it suitable for e.g.\ future $\beta$-asymmetry measurements using the low temperature nuclear orientation technique. 

\begin{table}[tb]%[H] add [H] placement to break table across pages
\centering
\caption{\label{tab:pips_depl}Simulated ratio of the $^{207}$Bi conversion electron peaks K$_{1063}$/K$_{569}$ for different depletion layer thickness of the PIPS detector. The $\chi^2_{red}$ value shown is calculated according to Equation~\ref{eq:chidef} when comparing the simulated spectra and the experimental spectrum taken at 77~K, for the energy region between 100 and 1100~keV. The statistical uncertainties on the simulated peak ratios are of the order of $1 \permil$. The experimental ratio is 3.78(11).}
\begin{ruledtabular}
 
\begin{tabular}{c c c}

Depletion (mm) & Peak ratios & $\chi^2_{red}$\\
\hline
1.39 & 3.71 & 23.1 \\
1.43 & 3.83 & 17.1 \\
1.47 & 3.94 & 13.5 \\
1.50 & 3.98 & 11.5 \\

\end{tabular}
\end{ruledtabular}
\end{table}

\subsection{Comparisons with Geant4}
Two spectrum measurements were performed with the PIPS detector (at 77~K), one with $^{60}$Co, the other with $^{207}$Bi, since the source geometry was best known for these two cases. For simulating the spectra for these isotopes, the vacuum chamber with support structures as well as detailed detector and source descriptions were again modeled in Geant4 version 9.5. In all cases the Standard physics list was used.

\subsubsection{\label{sec:60Co_pips}$^{60}$Co}
\begin{figure}
\includegraphics[width=0.5\textwidth]{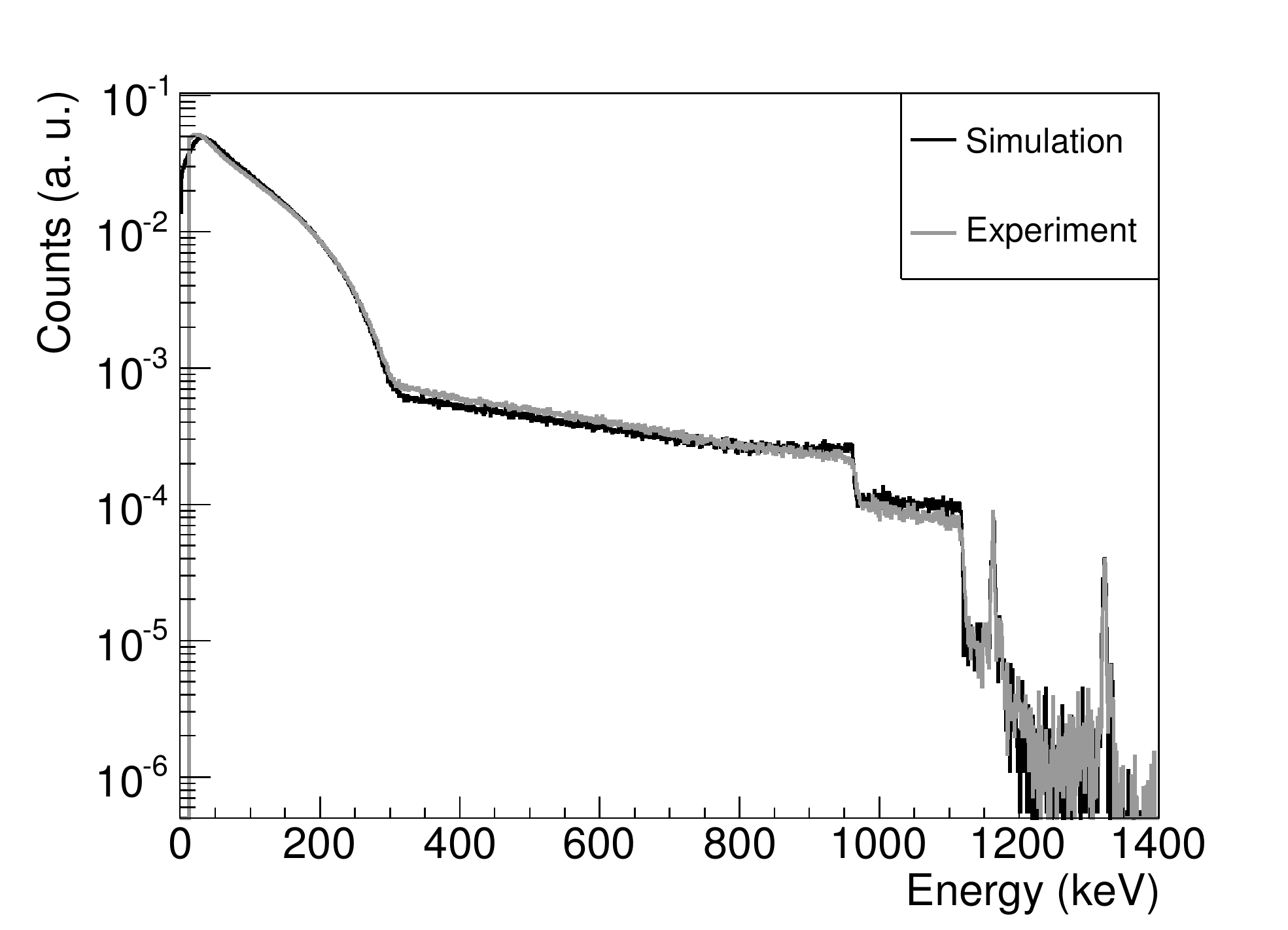}%
\caption{\label{fig:pips_co_simexp} Comparison between the simulated and measured spectra of $^{60}$Co for the PIPS detector. The $\beta$ endpoint energy is 317.9~keV. The two peaks at 1162 and 1320~keV are the conversion electrons of the two  $\gamma$-rays at 1173.2 and 1332.5~keV, respectively. The $\gamma$ peaks are estimated to be $\sim$20 times weaker than the conversion peaks. The simulation was performed with the Standard physics list and the Urban MSC model, while the simulation parameter values were set according to Section~\ref{sec:geant} with CFS fixed at 1~$\mu$m. Both spectra were normalized to the number of counts in the energy region of 150-300~keV.}
\end{figure}

A Si detector is suitable for precision $\beta$ decay measurements with $^{60}$Co since the low energy $\beta$-rays of this isotope will be fully stopped if the detector is at least 0.5~mm thick. The two gamma lines at 1.173 and 1.332~MeV will not create visible peaks, although their conversion electrons do appear in the spectrum obtained with the 1.5~mm thick Si detector. Figure~\ref{fig:pips_co_simexp} shows the experimental and the simulated spectrum of $^{60}$Co for the PIPS detector. The simulations were obtained using the Standard physics list and the Urban MSC model with the default values for the simulation parameters mentioned in Section~\ref{sec:geant} (i.e.\ CFS~=~1~$\mu$m, $F_R=0.04$, $F_G=2.5$ and Skin~=~3). 

\begin{figure}
\includegraphics[width=0.5\textwidth]{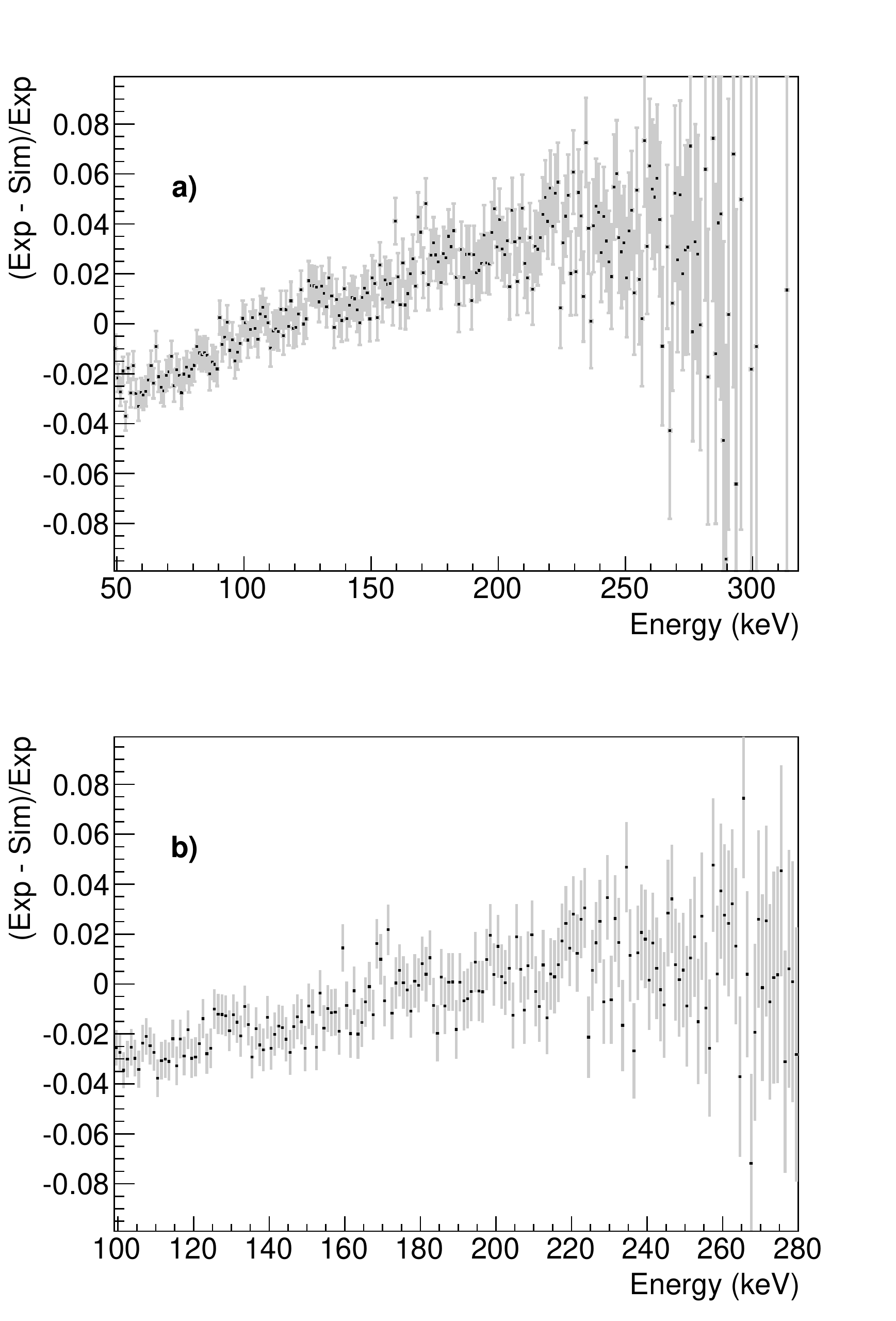}%
\caption{\label{fig:pips_co_pvcabs} Relative difference between simulated and experimental spectra of $^{60}$Co for the 1.5~mm PIPS detector. The experimental and simulated results were obtained by subtracting spectra with and without a 3~mm thick PVC absorber between the source and the detector. Simulations used the Single Scattering model, while the other parameters were set according to Section~\ref{sec:geant}, with CFS fixed at 1~$\mu$m. Panel a) was normalized in the region between 50 and 318~keV yielding $\chi^2_{red}=4.1$ for 268 degrees of freedom. Panel b) was normalized between 100 and 280~keV with $\chi^2_{red}=1.6$ for 180 degrees of freedom.}
\end{figure}

The clear difference in the region of the spectrum dominated by Compton events (i.e.\ above 320~keV) was further investigated by positioning a 3~mm thick plastic (PVC) absorber between the detector and the source in order to block the $\beta$ rays. Keeping in mind that the PVC absorber slightly increases the Compton background in both simulations and experiment, one can subtract the spectrum with absorber from the regular one by normalizing the spectra in the energy region above the $\beta$ decay endpoint (i.e.\ between 350 and 600~keV). This procedure reduces the influence of the Compton events to a second order effect, so that the difference between simulation and experiment is now below 5\% in the region from 50 to 318~keV and below 3\% when normalizing between 150 and 300~keV (see figure~\ref{fig:pips_co_pvcabs}). 

\subsubsection{Comparison of MSC models, physics lists and Geant4 parameters}
The fact that simulated and experimental spectra agree up to 3\% for the case of $^{60}$Co (Figure~\ref{fig:pips_co_pvcabs}) allows comparing the effect of different types of Geant4 simulation parameters on the simulated spectra. We therefore investigated the influence of the different physics lists and MSC models, as well as of the values of the different simulation parameters.

Performing simulations with the different physics lists it was found that, similar to the backscattering coefficients, the simulated spectra of $^{60}$Co were not significantly influenced by the choice of the physics list, i.e.\ similar $\chi^2_{red}$ values were obtained for both energy regions and for all three physics lists; see Table~\ref{tab:pips_60co_physlist_chi}. The amount of computing time required for physics lists other than the Standard one is found to be roughly two times larger. 

In simulations performed using the different MSC models the $\chi^2_{red}$ values listed in Table~\ref{tab:pips_60co_msc_chi} were obtained. The Single Scattering and the Goudsmit-Saunderson models clearly outperform the Urban MSC model, yielding a $\chi^2_{red}$ value that is up to about 50\% smaller when the larger energy region from 50 to 318~keV is considered. The Goudsmit-Saunderson model performs similar to the Single Scattering model, but because of the straggling in the backscattering coefficients (see Figure~\ref{fig:backscatt_MSC}) it can not be recommended. 

Comparing simulations performed with different CFS values the results that are summarized in Table~\ref{tab:pips_60co_cfs} were obtained. Keeping in mind that a CFS value of 1~mm is too large when simulating the performance of detectors which are only several mm thick, the best value for the CFS is found to be around 10~$\mu$m to 0.1~mm, with a preference for the smaller value in view of this size issue. 

Finally, simulations were performed for $F_R$~=~0.002 and for the default value of 0.04, resulting in $\chi^2_{red}$~=~6.3 and 7.75, respectively, for the energy region of 50 to 318~keV. This difference being less significant than the differences in $\chi^2_{red}$ obtained when varying the physics lists, the MSC models or the CFS value we suggest to keep the default value for $F_R$. The $F_G$ and Skin parameters were found not to influence significantly the simulated spectra. 

In order to investigate the accuracy of the electron processes within Geant4 in greater detail, experimental data for pure ground state to ground state $\beta$ transitions are required so that no Compton effect has to be considered. 

\begin{table}[h]%[H] add [H] placement to break table across pages
\centering
\caption{\label{tab:pips_60co_physlist_chi} $\chi^2_{red}$ values obtained by comparing experimental and simulated spectra of $^{60}$Co in two different energy regions, using the Standard (Std.), Penelope (Pen.) and Livermore (Liv.) physics lists.  The CFS was set to 1~$\mu$m, while the $F_R$, $F_G$ and Skin simulation parameters were set at their default values given in Section~\ref{sec:geant}.} 
\begin{ruledtabular}
\begin{tabular}{y c c c }

\multicolumn{1}{c}{Region (keV)}& Std. &  Pen. & Liv.\\
\hline
150-300 & 1.53 & 1.13 & 1.31  \\
50-318 & 7.75 & 7.14 & 7.97  \\
\end{tabular}
\end{ruledtabular}
\end{table}

\begin{table}[h]%[H] add [H] placement to break table across pages
\centering
\caption{\label{tab:pips_60co_msc_chi} $\chi^2_{red}$ values obtained by comparing experimental and simulated spectra of $^{60}$Co in two different energy regions, using the Standard physics list with the Urban, Goudsmit-Saunderson (G-S) and Single Scattering (SS) MSC models. The CFS was set to 1~$\mu$m, while the $F_R$, $F_G$ and Skin simulation parameters were set at their default values given in Section~\ref{sec:geant}.} 
\begin{ruledtabular}
 \begin{tabular}{y c c c }
\multicolumn{1}{r}{Region (keV)}& Urban &  G-S & SS \\
\hline
150 - 300 & 1.53 & 1.25 & 1.10\\
50 - 318 & 7.75 & 3.71 & 4.76\\
\end{tabular}
\end{ruledtabular}
\end{table}

\begin{table}[h]%[H] add [H] placement to break table across pages
\centering
\begin{ruledtabular}
\caption{\label{tab:pips_60co_cfs} $\chi^2_{red}$ values obtained by comparing experimental and simulated spectra of $^{60}$Co in two different energy regions, using the Standard physics list with the Urban MSC model, for values of the CFS parameter ranging from 1~$\mu$m to 1~mm. The $F_R$, $F_G$ and Skin simulation parameters were set at their default values given in Section~\ref{sec:geant}. }
\begin{tabular}{y c c c c}
\multicolumn{1}{c}{Region (keV)}& 1 $\mu$m & 10 $\mu$m & 0.1 mm & 1 mm \\
\hline
150-300 & 1.53 & 1.03 & 1.07 & 1.13  \\
50-318 & 7.75 & 4.83 & 3.09 & 3.28  \\
\end{tabular}
\end{ruledtabular}
\end{table}

\subsubsection{$^{207}$Bi}
\begin{figure}
\includegraphics[width=0.5\textwidth]{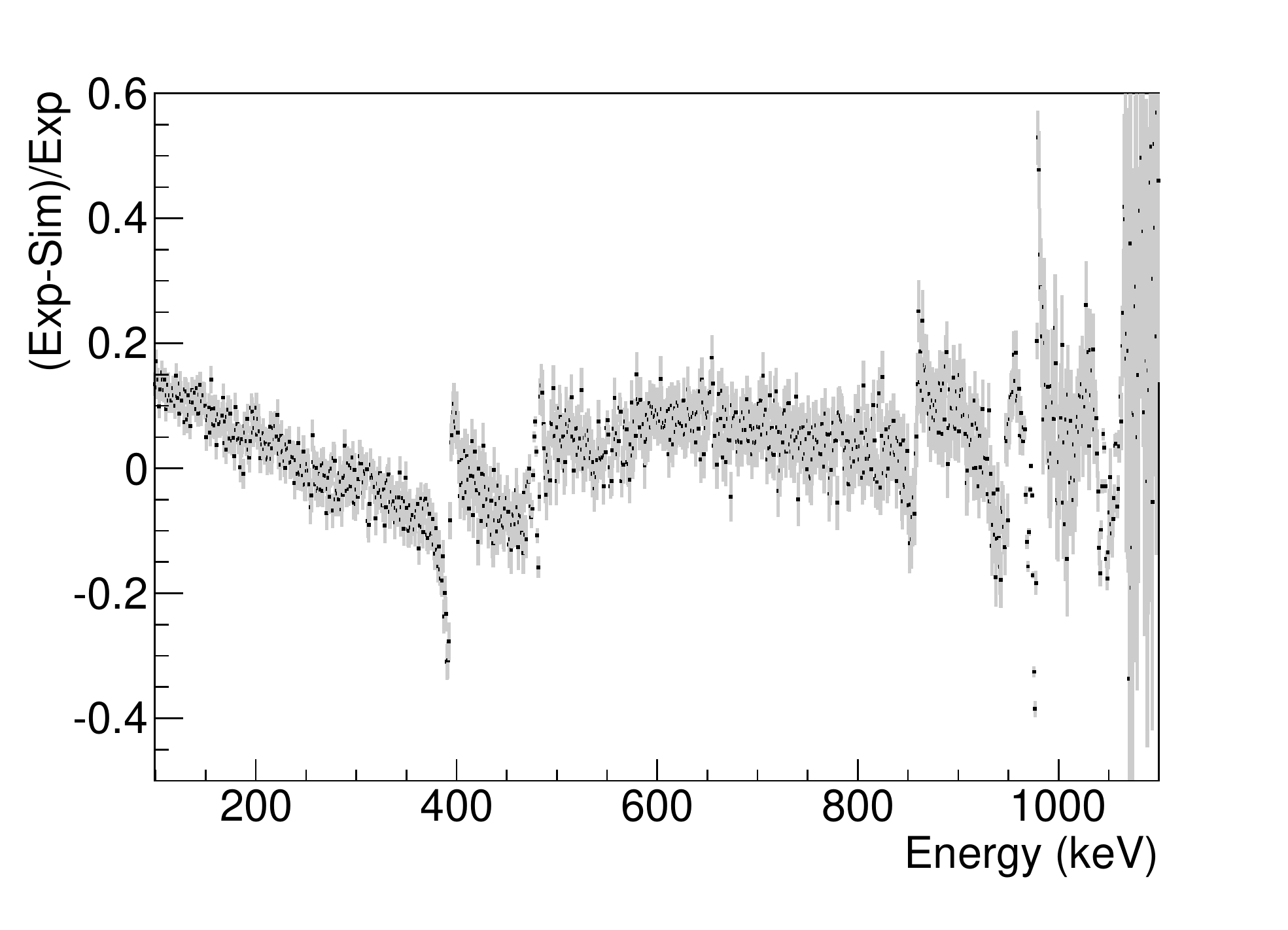}%
\caption{\label{fig:pips_bi_diffplot} Relative difference between the experimental and simulated spectrum for $^{207}$Bi. Simulation parameters were set according to Section~\ref{sec:geant}. The normalization region is from 100 to 1100~keV. The spikes visible at around 400 and 860~keV are the differences near the Compton edges; see text and Figure~\ref{fig:pips_bi_ComptonEdge} for details.}
\end{figure}

This isotope is the most demanding for Geant4, as was already discussed in Section~\ref{sec:hpge_comp}. Figure~\ref{fig:pips_bi_exp} shows the experimentally obtained spectrum, while Figure~\ref{fig:pips_bi_diffplot} shows the difference between simulation and experiment. In the region of the spectrum dominated by the X-rays (at 75 and 85~keV) the difference increases to 40\%, while in the higher energy region it is around 10\%. 

\begin{figure}
\includegraphics[width=0.5\textwidth]{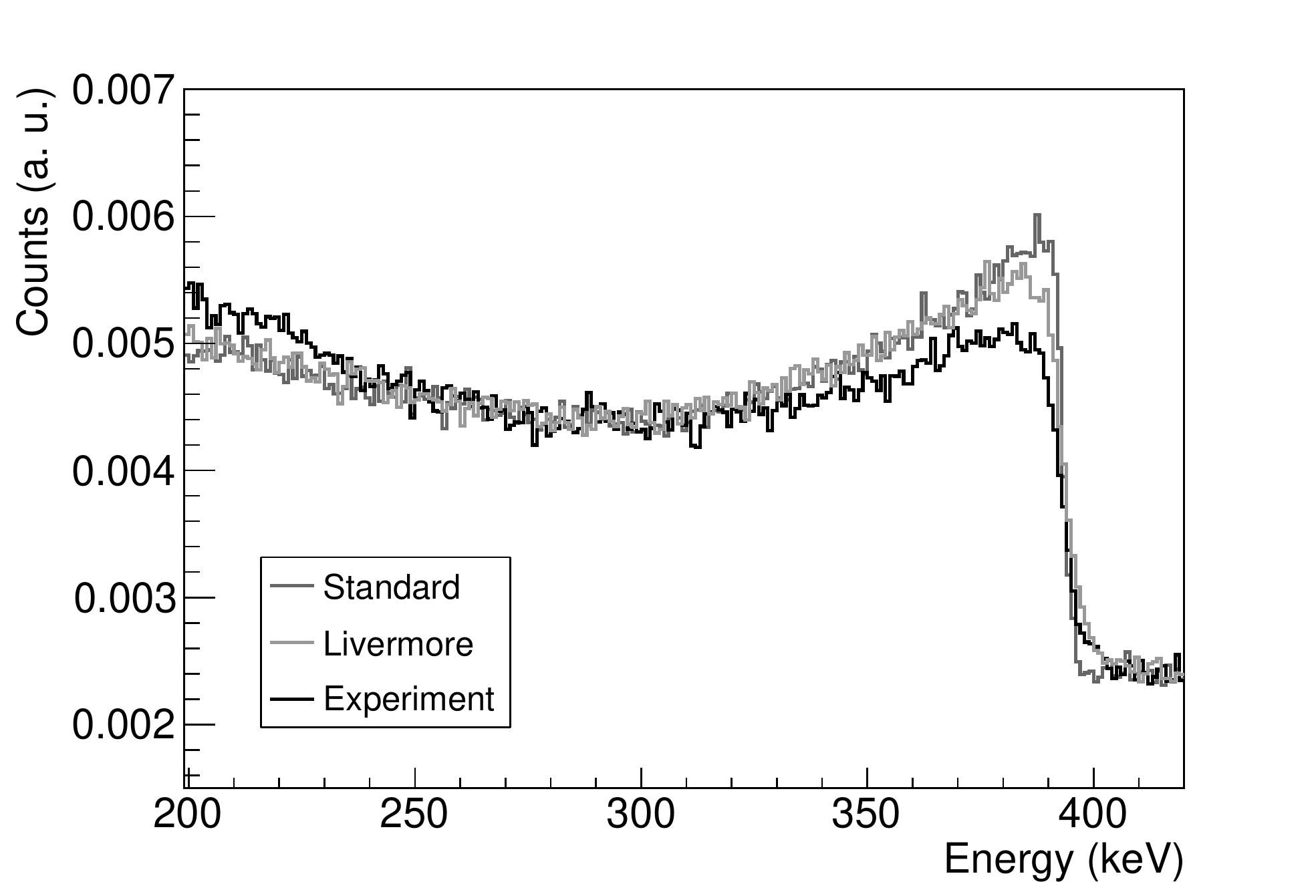}%
\caption{\label{fig:pips_bi_ComptonEdge} The Compton edge of the 569.7~keV $\gamma$ line of $^{207}$Bi, measured using the PIPS detector. The simulated spectra were obtained using the Standard and the Livermore physics list with the Urban MSC model.}
\end{figure}

The most significant difference is observed in the region of the Compton plateau of the 569.7~keV $\gamma$ line, i.e.\ the energy region up to about 400~keV. Further, the Compton edge in the simulated spectrum is much sharper and more intense by $\sim$15\%. Using different physics lists was found to yield slightly different results. E.g.\ the spectrum generated with the Livermore physics list displays a smoother Compton edge than the Standard one, in better agreement with the experimental spectrum (see Figure~\ref{fig:pips_bi_ComptonEdge}). However, the difference in the intensity of the Compton edge is still of the order of 10\%. The Penelope physics list was found to produce very similar results. 


\section{Conclusions and outlook}
The influence of the various Geant4 physics lists, MSC models and parameters on simulated electron backscattering coefficients from Si for energies in the range of nuclear $\beta$ decay was investigated. It was found that for the energy region of typical low energy experiments in neutron and nuclear $\beta$ decay the usage of the low energy physics lists - Livermore and Penelope - is not absolutely required. Best agreement between the simulated results and experimental data is found for the Single Scattering model, although the other MSC models are also within the (still rather large) experimental uncertainty. The default value for CFS, which is 1~mm, should be lowered for geometrical reasons and our recommended value is around 10~$\mu$m. Our recommended value for the $F_R$ parameter is in the range between 0.01 and 0.002. High precision experimental data on backscattering, but also on transmission through thin foils, in the energy range of 100 and 1000~keV would be very useful for further investigation. Such a project is currently being prepared using a $\beta$ spectrometer which combines an energy sensitive detector and a multi-wire drift chamber similar to the one described in Refs.~\cite{Ban2006, Lojek2009}.

Simulations of the response of semiconductor particle detectors (3~mm thick HPGe and 1.5~mm thick PIPS) were also compared to experimental data. A general good agreement was found for electron processes, while for $\gamma$ processes significant differences were observed in the region of the Compton edge. The overall worse agreement for HPGe detectors is not surprising due to the higher $Z$ value of Ge, since both electron and $\gamma$ processes are $Z$ dependent. 

The observed good accuracy that was found for the electron processes in Geant4 now allows for direct precision measurements of the $\beta$ spectrum shape. For the case of a pure $\beta$ emitter, with no $\gamma$ rays in its decay, the dominant systematic effect remaining is the backscattering from the energy sensitive detector. Combining then the detector with a system that identifies backscattered events (such as the multi-wire drift chamber mentioned above \cite{Ban2006,Lojek2009}) enables performing high precision measurements of the $\beta$ spectrum shape. Such experiments are currently being prepared. These would allow to address the Fierz interference term that is sensitive to scalar and tensor type components in the weak interaction \cite{Jackson1957}, and to study the effect of the so-called recoil terms \cite{Holstein1974} in nuclear $\beta$ decay.

\section*{Acknowledgments}
This work was supported by the Fund for Scientific Research Flanders (FWO), Projects GOA/2004/03 and GOA/2010/10 of the K.\ U.\ Leuven, the Interuniversity Attraction Poles Programme, Belgian State Belgian Science Policy (BriX network P6/23),
and Grant LA08015 of the Ministry of Education of the Czech Republic.

\bibliography{master}

\end{document}